\documentclass[3p,times]{elsarticle}

\usepackage{ecrc}
\newcommand{\bald}[1]{{\bf #1}}
\volume{00}

\firstpage{1}

\journalname{Nuclear Physics A}

\runauth{B. W. Zhang}

\jid{nupha}

\usepackage{amssymb}
 \usepackage{lineno}

\usepackage[figuresright]{rotating}

\pdfoutput=1
\begin{document}

\begin{frontmatter}



\dochead{}

\title{Full jet tomography of high-energy nuclear collisions}


\author{Ben-Wei Zhang}

\address{Institute of Particle Physics, Central China Normal University, Wuhan 430079, China}
\address{Key Laboratory of Quark $\&$ Lepton Physics (Central China Normal
University), Ministry of Education, China}

\begin{abstract}
Parton energy loss in the hot QCD medium will manifest itself not only in leading hadron spectra but also in
reconstructed jet productions in high-energy nucleus-nucleus collisions. In this paper we report
 on recent theoretical efforts in studying full jet observables in relativistic heavy-ion collisions by discussing
 the modifications of jet shapes, inclusive jet cross section and the vector boson accompanied jet production
 in the presence of the QGP-induced jet quenching.

\end{abstract}


\end{frontmatter}

 %%\linenumbers

\section{ Introduction}
\label{sec:intro}

In high-energy nucleus-nucleus collisions a deconfined matter, the quark-gluon plasma (QGP)
is to be formed and when an energetic parton produced by the initial hard scattering propagates
in this hot and dense QCD medium, it will interact with other partons in the hot nuclear medium
and be quenched substantially\cite{Wang:1991xy,GVWZ,BDMPS-Z-ASW,GLV,HT}.  This parton energy loss mechanism,
or jet quenching,
has been proposed to be an excellent probe
of the QGP and predicted the suppression of single hardon productions and the disappearance of
away-side dihadron correlations in heavy-ion collisions (HIC), which  later have been
confirmed by the experimental measurements at RHIC  and given compelling evidence of
the existence of a new kind of matter in high-energy nuclear collisions at RHIC \cite{RHIC}.  
Not surprisingly, the finding of jet quenching in heavy-ion collisions  has been regarded as one of the most important
discoveries made at RHIC and initiated many theoretical explorations
and very precise experimental measurements \cite{d'Enterria:2009am}.

So far the dominant experimental measurements of jet quenching is about the production of one or two hadrons 
with a large transverse momentum,  which are only the leading fragments of a jet. The parton energy loss
mechanism will not only manifest itself in leading particle productions, but more interestingly 
in full jet observables.  Recently in heavy-ion community a large amount of effort has been invested in measuring
reconstructed jets in high-energy nuclear collisions for the first time\cite{JetExp,Jetlhc,JetExp@HP2010}, and theories to address
the novel features of full jet tomography in relativistic heavy-ion collisions have emerged. In this article
we review our theoretical studies on jet observables of high-energy nucleus-nucleus collisions in
perturbative Quantum Chromodynamics (pQCD) by 
focusing on the intra-jet energy flow, the inclusive jet cross section and $Z^0/\gamma^*$-tagged jet production 
\cite{VWZ,Vitev:2009rd,Neufeld:2010fj} as well as
possible non-perturbative effects in jet productions.

\section{ Jet shapes in HIC}
\label{sec:shapes}
One of the most common obesevables of resolving the internal jet structure
 is the jet shape, which describes the energy distribution
in a jet as \cite{Seymour:1997kj}:
\begin{equation}
 \Psi_{\rm {int}}(r;R) = \frac{\sum_i (E_T)_i \Theta (r-R_{i, \rm {jet}})}
{\sum_i  (E_T)_i  \Theta(R-R_{i,\rm {jet}})} \, .
\end{equation}
Here $r,R$ are Lorentz-invariant opening angles, $R_{i,{\rm jet}} =                       
\sqrt{(y_i-y_{\rm {jet}})^2 + (\phi_i-\phi_{\rm {jet}})^2}$ gives the distance
between a parton $i$ and the jet in the space of rapidity $y$ and azimuthal $\phi$.

In leading order we can derive the differential jet shape as \cite{VWZ}
\begin{equation}
   \psi(r;R) = \frac{ d  \Psi_{\rm {int}}(r;R) }{ dr } =
\sum_b \frac{\alpha_s}{ 2 \pi }
\frac{2}{r} \int_{z_{min}}^{1-Z} dz  \,  z P_{a \rightarrow bc}(z),
\label{kernel}
\end{equation}
where
\begin{eqnarray}
  Z &=& \max \left\{ z_{min},   \frac{r}{r+R} \right\} \;  \mbox{ if }
\;
r < (R_{sep} - 1)R  \;,  \nonumber \\
   Z  &=& \max \left\{ z_{min},  \frac{r}{R_{sep}R} \right\} \;
\mbox{ if } \;
r > (R_{sep} - 1)R  \; . \nonumber
\label{Zpar}
\end{eqnarray}
Here $1 \leq R_{sep} \leq 2$ is introduced 
to take into account features of experimental cone algorithms, employed to improve 
infrared safety. In Eq.~(\ref{kernel})  $ r = (1-z) \rho $ is related to the opening
angle $\rho$ between the final-state partons, $P_{a \rightarrow bc}(z)$ is the
splitting kernel in DGLAP evolution equation, $z$ is the momentum
fraction of the radiated parton relative to the parent parton,
$z_{min} = E_T^{min}/E_T$ with $E_T$ representing the
transverse momentum of a jet, and $E_T^{min}$
the minimum transverse momentum of a parton in this jet .

%%%%%%%***********Figure for theoretical calculations of jet Shape **************
%\vspace{0.3in}
\begin{figure}[t]
\begin{center}
%\vspace*{-1.in}
%\hspace{-0.4cm}
\includegraphics[width=2.8in,height=2.9in,angle=0]{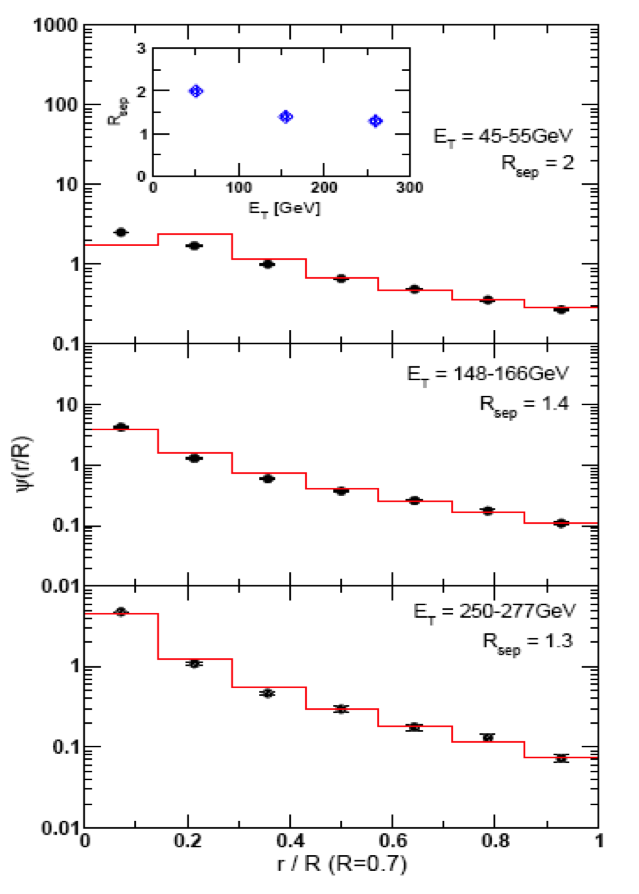} 
\hspace{0.6cm}
\includegraphics[width=2.8in,height=2.9in,angle=0]{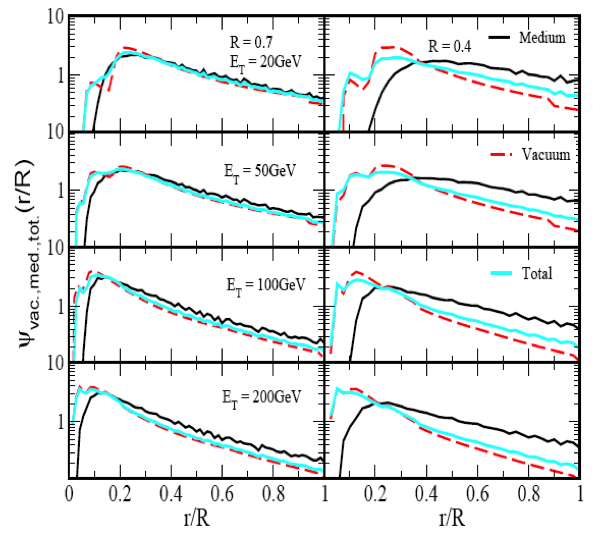} \vspace*{-.1in}
\caption{ Comparison of numerical results
from our theoretical calculation to experimental data on
differential jet shapes at $\sqrt{s}=1960$~GeV by
CDF II~\cite{Acosta:2005ix} (left panel).  Numerical simulation of jet shapes in vacuum
, medium-induced jet shapes, and total jet shapes in medium for
 Pb+Pb with $\sqrt{s_{NN}}=5.5$~TeV at LHC (right panel). }
\label{fig:shape-1}
\end{center}
\end{figure}
%%%%%%%%*************************************************************

Carrying out the integration of $z$ we get the jet shapes at LO.  
For instance, the jet shape for a quark gives:
\begin{eqnarray}
  \psi_q(r) = \frac{C_F \alpha_s}{2\pi} \frac{2}{r}  \left(
  2 \log \frac{1-z_{min}}{Z}  \right. 
 \left. - \frac{3}{2} \left[ (1-Z)^2 -z_{min}^2 \right]  \right)  \; ,
\label{psiLO1} 
\end{eqnarray}

%%%%%%%***********Figure for theoretical calculations of jet Shape **************
%\vspace{0.3in}
\begin{figure}[t]
\begin{center}
%\vspace*{-1.in}
%\hspace{-0.4cm}
\includegraphics[width=2.8in,height=2.2in,angle=0]{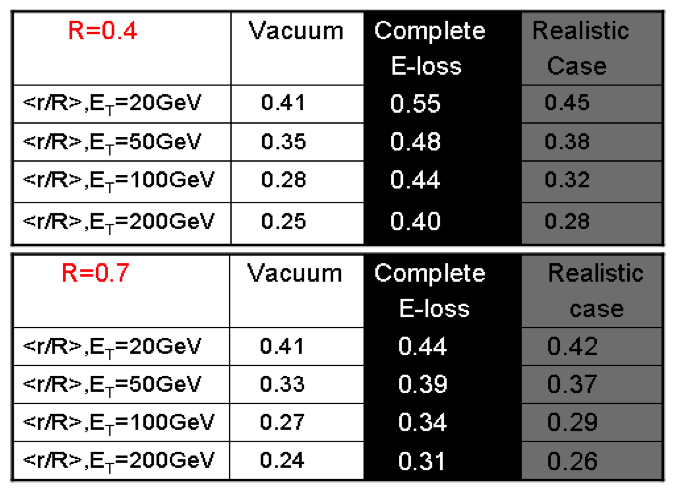} 
\hspace{0.6cm}
\includegraphics[width=2.8in,height=2.2in,angle=0]{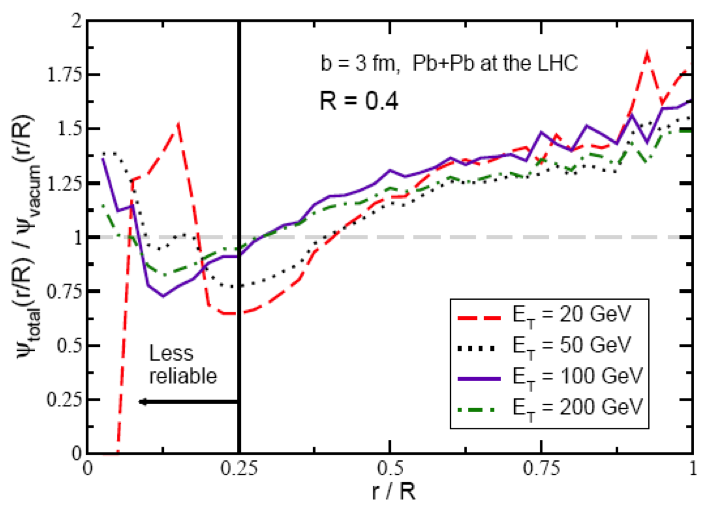} \vspace*{-.1in}
\caption{ The mean relative jet radii $\langle r/R \rangle$
in the vacuum, with complete energy loss, and  in the QGP medium for
 $ \sqrt{s}=5.5$~TeV central Pb+Pb collisions at the LHC (left panel);
 The ratios of total jet shape in heavy-ion collisions at LHC to the jet shape in the vacuum (right panel). }
\label{fig:shape-2}
\end{center}
\end{figure}
%%%%%%%%*************************************************************

Eq.~(\ref{psiLO1}) is collinear divergent when $r \rightarrow 0$. To get rid of this divergence,
we need to make Sudakov resummation. Also we could improve the prediction power of our 
analysis by estimating the non-perturbative effect and include contributions of the initial-state radiation and
corresponding Sudakov resummation \cite{Seymour:1997kj}. 
Taking into account all these contributions, we do numerical simulation and show the
comparison of
the theoretical results for jet shapes in $p+{\bar p}$ collisions with $\sqrt{s_{NN}}=1960$~GeV
at Tevatron with CDF II data in Fig.~\ref{fig:shape-1}. It could be seen that the theoretical model describes CDF II data
very well, especially jet shapes away from $r=0$ \cite{VWZ}.

In a hot and dense QCD medium,  a propagating fast parton may lose a large amount of energy
by medium-induced gluon radiation, and some lost energy carried away by radiated gluon may be 
recaptured by the jet  and thus gives additional contributions
to jet shapes.  In our studies, we calculate the radiated gluon angular distribution in medium within the GLV formalism 
~\cite{GLV} and compute the medium-induced jet shapes  as well as the in-medium jet cross sections 
(see Sec.\ref{sec:jcs} and Sec.\ref{sec:Z-jet} ) using full numerical evaluation  of
the medium-induced contribution to the parton showers.  In the model energetic inclusive jet production 
and tagged jet production are rare  processes that follow binary collision  scaling $ \sim d^2N_{\rm bin}/d^2{\bf x} $ 
with $d^2{\bf x}$ representing the area in the transverse plane; in contrast, the medium is distributed according to the number of participants density $ \sim d^2N_{\rm part}/d^2{\bf x}$ and 
the longitudinal Bjorken expansion of the QGP incorporated.

The total jet shapes in HIC then is given by\cite{VWZ}: 
\begin{eqnarray}
\psi_{\rm tot.}\left(\frac{r}{R}\right) \propto  \int_{\epsilon=0}^1
d\epsilon \; \sum_{q,g} \frac{P_{q,g}(\epsilon,E)  } 
{ (1 - (1-f_{q,g}) \cdot \epsilon)^3} \,   \Bigg[ (1- \epsilon) \;
\psi_{\rm vac.}^{q,g}\left(\frac{r}{R};E^\prime \right) + f_{q,g}\cdot \epsilon \;
\psi_{\rm med.}^{q,g}\left(\frac{r}{R};E^\prime \right) \Bigg] \; , \qquad
\label{psitotmed}
\end{eqnarray}
where $(1-f_{q,g}) \cdot \epsilon$  gives the fraction of the 
energy of the parent parton falling outside of jet area with 
a radius $R$.
In the right panel of  Fig.~\ref{fig:shape-1} we illustrate  jet shapes in vacuum $\psi_{\rm vac.}(r/R)$, medium-induced
jet shapes $\psi_{\rm med.}(r/R)$, and total jet shapes in HIC $\psi_{\rm tot.}(r/R)$ in Pb+Pb collisions 
with $\sqrt{s_{NN}}=5.5$~TeV. It is observed that
though medium-induced jet shapes are quite distinct from jet shapes in vacuum, the difference between
jet shapes in p+p collisions and that in Pb+Pb is not very large.  The underlining reason for this surprising result is that although
medium-induced gluon radiation produces a broader $\psi_{\rm med.}(r/R)$,
this effect is offset by the fact that jets lose a finite
amount of their energy, which will shift jet shapes in vacuum to higher $E_T$ 
with steeper profile.  The left panel of Fig.~\ref{fig:shape-2} shows the mean relative jet
radii~$ \langle r/R \rangle $ in the vacuum and in the QGP medium
created at the LHC for  two different cone selections $R=0.4$ and
$R=0.7$.  We can find that the broadening of radii  $ \langle r/R \rangle $ of jet shapes in HIC is 
modest, which implies only a fraction of energy of jets lost in the hot medium and the QGP is
rather 'gray' instead of 'black' \cite{VWZ}. The deviation of $\psi_{\rm tot.}(r/R)$ from $\psi_{\rm vac.}(r/R)$
is pronounced in the tail of jet shapes for small cone size $R=0.4$, as illustrated 
in the right panel of  Fig.~\ref{fig:shape-2}.

\section{ Inclusive jet cross sections in HIC}
\label{sec:jcs}

%%%%%%%***********Figure for theoretical calculations of jet CS  at RHIC**************
%\vspace{0.3in}
\begin{figure}[t]
\begin{center}
%\vspace*{-1.in}
\hspace{-0.4cm}
\includegraphics[width=2.8in,height=2.1in,angle=0]{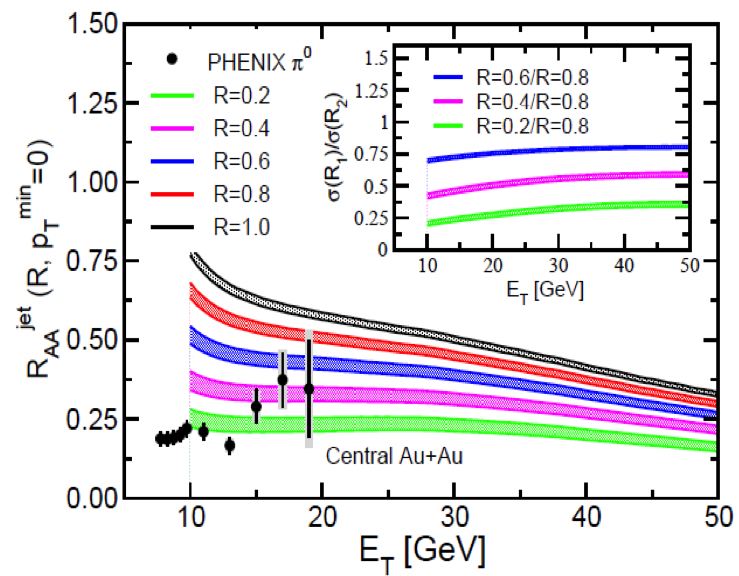} 
\hspace{0.6cm}
\includegraphics[width=2.8in,height=2.0in,angle=0]{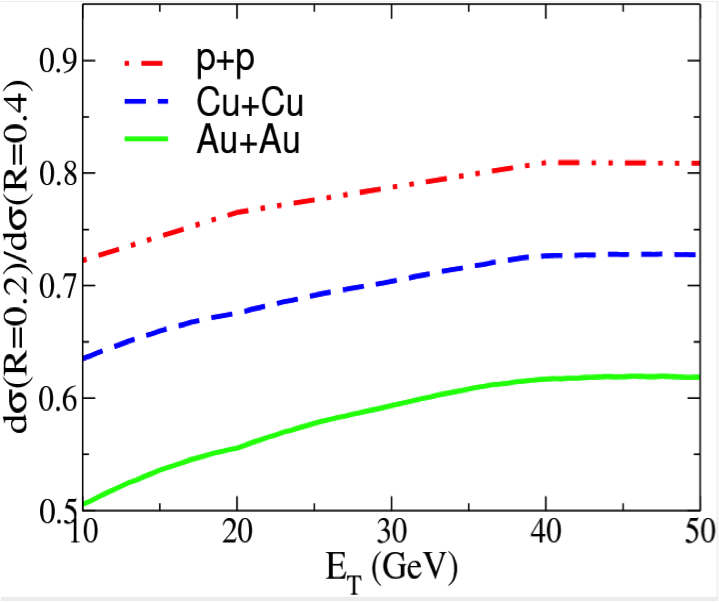} \vspace*{-.1in}
\caption{ Transverse energy dependent nuclear modification factor $R_{AA}^{\rm jet}$
for different cone radii $R$ at NLO in $b=3$~fm Au+Au collisions at $\sqrt{s_{NN}}=200$~GeV (left panel). 
The ratio of jet yields at $R=0.2$ and $R=0.4$ in p+p, central Cu+Cu and
Central Au+Au collisions  with $\sqrt{s_{NN}}=200$~GeV at RHIC (right panel). }
\label{fig:jet@NLO}
\end{center}
\end{figure}
%%%%%%%%*************************************************************

Parton energy loss in the QGP not only changes the energy distribution in a jet, but
also alters the inclusive jet spectrum in high-energy nucleus-nucleus reactions. A first
study of inclusive jet productions in HIC was carried out in the leading order (LO)
 \cite{VWZ}, and recently
 to take full advantage of jet physics in reactions with ultra-relativistic nuclei
a next-to-leading (NLO) calculations at 
${\cal O}(\alpha_s^3)$  have been made~\cite{Vitev:2009rd}.

In hadron-hadron collisions 
the inclusive jet cross section at NLO can be given as ~\cite{EKS}:
\begin{eqnarray}
\frac{d\sigma^{\rm jet}} {dE_Tdy}=\frac{1}{2!}
\int d\{E_T,y,\phi\}_2 
\frac{ d\sigma[2\rightarrow 2]}{d\{E_T,y,\phi\}_2}  S_2 (\{E_T,y,\phi\}_2) +\frac{1}{3!}
\int d\{E_T,y,\phi\}_3   \frac{ d\sigma[2\rightarrow 3]}{d\{E_T,y,\phi\}_3 }
S_3 (\{E_T,y,\phi\}_3 ) \; ,\qquad
\label{eq:CS_NLO}
\end{eqnarray}
where $E_{T\,i},y_i,\phi_i$ are the transverse energy, rapidity, and azimuthal angle
of the i-th particle ($i=1,2,3$), respectively, and
$\sigma[2\rightarrow 2]$, $\sigma[2\rightarrow 3]$ stand for the partonic cross 
sections. In Eq.~(\ref{eq:CS_NLO}) $S_2$, $S_3$ give the space constraints and
contain the information of the jet-finding algorithm.   At leading order, a jet is 
equivalent to a parton and we always have $S_2 = \sum_{i=1}^{2} S(i)= 
\sum_{i=1}^{2} \delta(E_{T_i} - E_T)\delta(y_i -y)$. At NLO due to higher
order corrections $S_3$ has two contributions: one coming from the possibility that each jet
still contains one parton; another resulting from the chance that a jet contains two partons 
imposed by the jet finding algorithm \cite{Campbell:2006wx}. 
Numerical simulation  shows that the inclusive jet spectrum in p+p
collisions with $\sqrt{s_{NN}}=200$~GeV at RHIC is described very well by a NLO code of jet production in hadronic
collisions \cite{EKS}, which will later be extended to study inclusive jet productions at NLO in HIC.

In the same spirit of Eq.~(\ref{psitotmed}) for jet shapes in HIC, we derive the related medium-modified jet cross section
 as follows~\cite{Vitev:2009rd}:
\begin{eqnarray}
\frac{1}{\langle  T_{AA} \rangle \sigma^{\rm geo}_{AA}  } 
\frac{d\sigma^{AA}(R)}{d^2E_Tdy} = \int_{\epsilon=0}^1
d\epsilon \; \sum_{q,g}  P_{q,g}(\epsilon,E) 
\frac{1}{ (1 - (1-f_{q,g}) \cdot \epsilon)^2} 
 \frac{d\sigma^{\rm CNM,NLO}_{q,g}(R)} {d^2E^\prime_Tdy} \; . \qquad
\label{eq:JCS-AA}
\end{eqnarray}
Here $T_{AA}$ is the nuclear overlap function, and $  \sigma^{\rm geo}_{AA}   $ the geometrical A+A cross section
\cite{d'Enterria:2003qs}.
The measured cross section is then a probabilistic superposition of 
the  cross sections of protojets of initially larger energy
$E^\prime_T = E_T / (1 - (1-f_{q,g})\cdot \epsilon)$.

%%%%%%%***********Figure for jet CS measurement at RHIC**************
%\vspace{0.3in}
\begin{figure}[t]
\begin{center}
%\vspace*{-1.in}
\includegraphics[width=2.8in,height=2.2in,angle=0]{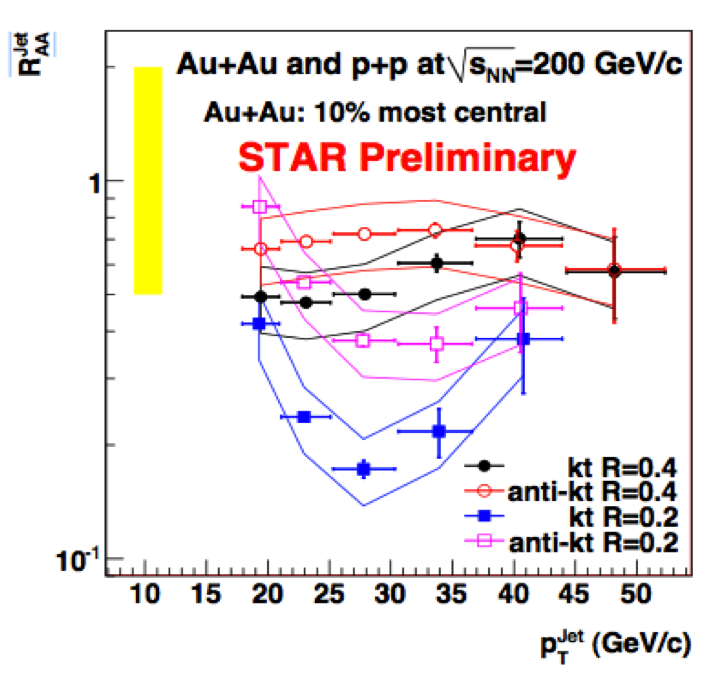} 
\hspace{0.6cm}
\includegraphics[width=2.8in,height=2.2in,angle=0]{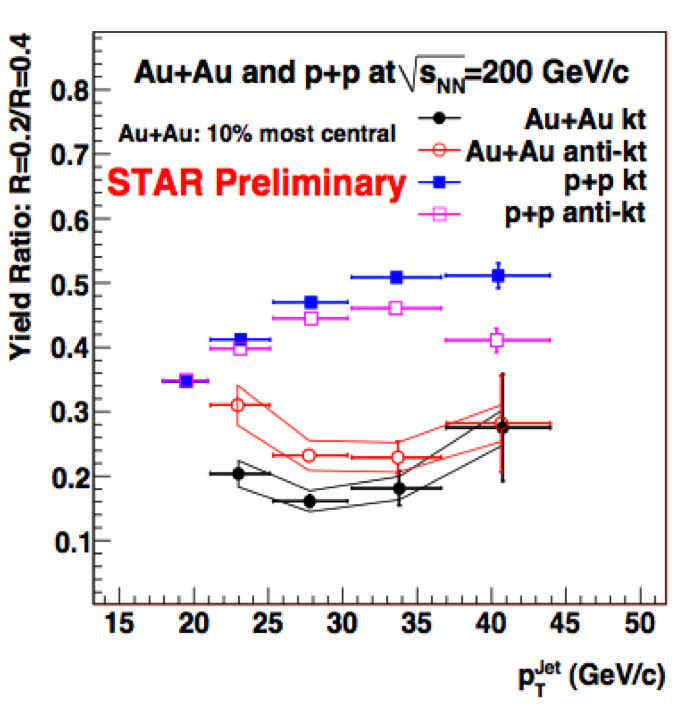} \vspace*{-.1in}
\caption{ Experimental measurements of full jet production in HIC with $\sqrt{s_{NN}}=200$~GeV 
at RHIC \cite{JetExp,JetExp@HP2010}:  the nuclear modification factor of jet production in Au+Au collisions 
at different jet radii (left panel); ratios of jet cross sections with $R=0.2$ to that with $R=0.4$ in
p+p and Au+Au collisions (right panel). }
\label{fig:jet@RHIC}
\end{center}
\end{figure}
%%%%%%%%*************************************************************

Now we are ready to compute the nuclear modification factor of jet cross section 
$R_{AA}^{\rm jet}$ defined as the ratio of jet cross section in HIC to that in p+p rescaled by the number
of binary collisions.   The numerical results for
the nuclear modification factor of inclusive jets $R_{AA}^{\rm jet}$ in central Au+Au
with $\sqrt{s_{NN}}=200$~GeV at RHIC are presented 
in the left panel of Fig.~\ref{fig:jet@NLO}~\cite{Vitev:2009rd}, where the band of theoretical calculations shows a calculation for a  
$\sim 20\%$ increase in the rate of parton energy loss  relative 
to our default simulation.  A continuous variation of
 $R_{AA}^{\rm jet}$  with the jet radius $R$  is found, which demonstrates a unique feature of jet production as compared to
the single curve of $R_{AA}$ for pion production in HIC. In the calculations several cold nuclear matter (CNM) effects
 such as nuclear shadowing, Cronin effect and EMC effect have been taken into account 
 \cite{Vitev:2008vk,Neufeld:2010dz}.  We observe that
 for $R \leq 0.2$ the quenching of jets approximates the
already observed suppression in the production  rate of inclusive high-$p_T$ particles.
In our theoretical calculation CNM effects contribute significantly
to the observed attenuation at large $E_T$. We could suppress the effects of CNM by taking the ratios of jet cross 
sections at different cone radii, which are plotted in the inserts of  the left panel of Fig.~\ref{fig:jet@NLO}.

To see the distinction of jet productions in different systems more clearly we plot the ratio 
$\frac{d\sigma^{AA}(R=0.2)}{dE_Tdy}/\frac{d\sigma^{AA}(R=0.4)}{dE_Tdy}$
for p+p collisions, central Cu+Cu and central Au+Au collisions with $\sqrt{s_{NN}}=200$~GeV
at RHIC in the right panel of Fig.~\ref{fig:jet@NLO} . It shows that the
ratio decreases as the system becomes larger, and the ratio in Au+Au at RHIC
is smaller than those in p+p and Cu+Cu collision at the same colliding energy, which
implies in HIC more contributions of jet yields  come from large angular radiation
 as compared to p+p collisions.

Recently PHENIX and STAR have measured reconstructed jets
in high-energy nucleus-nucleus collisions at RHIC and some preliminary results 
\cite{JetExp,JetExp@HP2010}  are illustrated in Fig.~\ref{fig:jet@RHIC}.  From the
left panel of Fig.~\ref{fig:jet@RHIC} one can see that measured $R_{AA}^{\rm jet}$ changes
with the jet radius and at smaller radius the nuclear modification factor of jets is smaller, which
agrees with our theoretical prediction \cite{VWZ,Vitev:2009rd} shown in the left panel of
Fig.~\ref{fig:jet@NLO}. Moreover the experimental measurements also demonstrate the
ratio of jet yields at different radii in Au+Au is smaller than the one in p+p (see the right panel
of Fig.~\ref{fig:jet@RHIC}) which confirms the
interesting feature shown  the right panel of Fig.~\ref{fig:jet@NLO} with our theoretical study.
We notice that though our theoretical predictions of inclusive jets describe very well the
overall trends of the jet measurements at RHIC, some deviations exist if
we compare the results in Fig.~\ref{fig:jet@NLO}  with the data in Fig.~\ref{fig:jet@RHIC}.
To confront the theoretical studies with the experimental data on a more solid base, 
we should include the correction from non-perturbative effects of jet productions 
\cite{Campbell:2006wx,Olness:2009qd,Dasgupta:2007wa}
in our theoretical investigation (please see Sec.\ref{sec:NP} 
for more discussions), and precise experimental measurements of jets in HIC with  large
statistics will also be needed.

\section{Tagged jet productions in HIC}
\label{sec:Z-jet}

%%%%%%%***********Figure for Z tagged jet at LHC**************
%\vspace{0.3in}
\begin{figure}[t]
\begin{center}
%\vspace*{-1.in}
\includegraphics[width=2.8in,height=2.2in,angle=0]{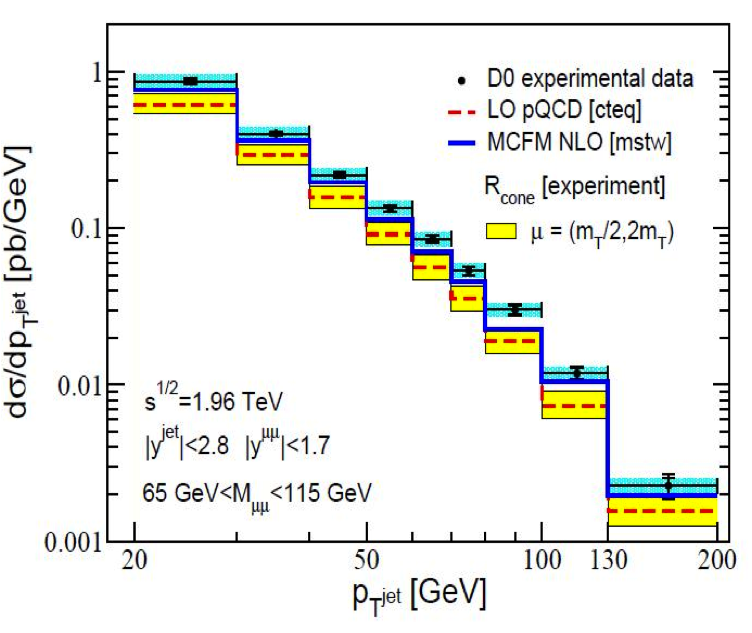} 
\hspace{0.6cm}
\includegraphics[width=2.8in,height=2.2in,angle=0]{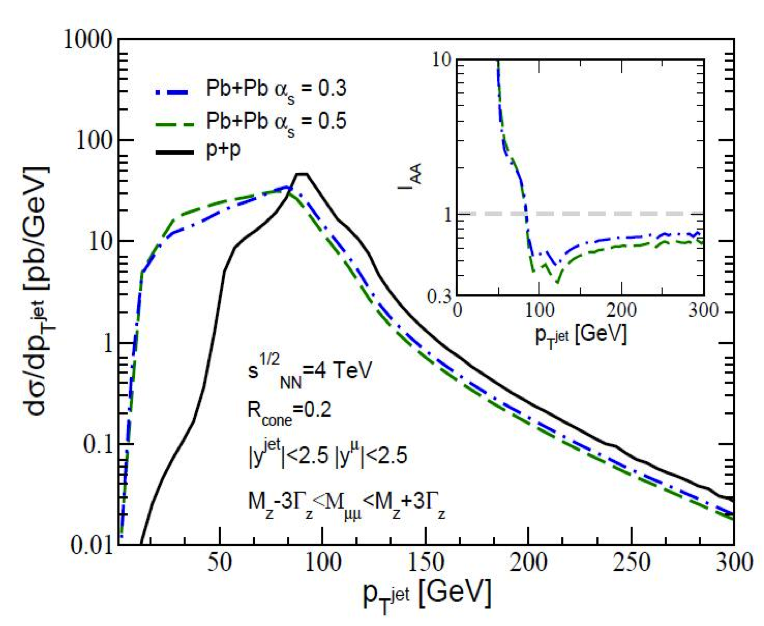} \vspace*{-.1in}
\caption{ Comparison of the theoretical calculation of $Z^0$/$\gamma^*$ tagged jet production in LO and NLO with MCFM to 
experimental results from the Fermi Lab Tevatron Collider \cite{Abazov:2008ez} for the cross section of jets associated with $Z^0$/$\gamma^*\rightarrow \mu^++\mu^-$ in $p+\bar{p}$ collisions at $\sqrt{s} = 1.96$ TeV (left panel).
The NLO $p_T$-differential cross section per nucleon pair for jets tagged with
$Z$/$\gamma^*\rightarrow \mu^++\mu^-$ in p+p and central Pb+Pb reactions when
$92.5 \; {\rm GeV} < p_T^{\rm jet} < 112.5\; {\rm GeV} $ (right panel).
}
\label{fig:Zjet}
\end{center}
\end{figure}
%%%%%%%%*************************************************************

A natural way to extend the study of the inclusive jet production in high-energy nucleus-nucleus reactions is to
consider tagged jet productions in HIC, especially the jet production accompanied with a vector boson ( $\gamma$
or $Z^0$ ) in nuclear collisions. In the leading order, the vector boson and the accompanied jet are produced back-to-back 
 in the center-of-mass frame 
and the transverse momentum of the vector boson $E_T^{\rm VB}$ is the same as that of the tagged 
jet $E_T^{\rm jet}$ before they travel in the QGP. 
Because the vector boson doesn't interact with the medium strongly its transverse momentum of the vector boson is intact.
By measuring $E_T^{\rm VB}$ and $E_T^{\rm jet}$ in the final stage of heavy-ion collisions we may know exactly the
energy loss of the jet by assessing $ E_T^{\rm VB} - E_T^{\rm jet}$.  Therefore the vector boson + jet production has been considered 
as an optimal channel to pinpoint the parton energy loss in the QGP and reveal the detailed information of  
the hot and dense QCD medium formed in relativistic HIC. However,  higher order corrections will break the 
momenta balance
between tagged jet and the vector boson due to additional radiation. Here we will review our
 recent study of $Z^0$/$\gamma^* + $jet production in Pb+Pb reactions at NLO and for a more detailed discussion we
 refer to \cite{Neufeld:2010fj,Neufeld:HP2010}.

For the baseline evaluation of tagged jet cross sections at NLO in hadron-hadron reactions we use the Monte Carlo for FeMtobarn  (MCFM) numerical code~\cite{Campbell:2002tg}. The confrontation of theoretical computation of tagged jet at MCFM NLO against
the experiment measurement for $p+{\bar p}$ collisions at Tevatron in the left panel of Fig.~\ref{fig:Zjet} 
attests to the validity of pQCD calculations. 
In heavy-ion collisions the medium-induced parton energy loss will modify the tagged jet spectrum and  the quenched jet cross section is given as follows:
\begin{eqnarray}
\frac{d\sigma^{AA} }{d^2 \bald{p}_{(Z)} d^2 \bald{p}_Q} = 
\sum_{q,g} \int d \epsilon \; P_{q,g}(\epsilon) \frac{1}{[1 - (1-f_{q,g}(\omega_{\min},R))\epsilon]^2} 
 \frac{d\sigma^{q,g}}{d^2 \bald{p}_{(Z)} d^2 \bald{p}_{\rm(jet)}}  \left(\frac{\bald{p}_{Q}}
{[1 - (1-f_{q,g})\epsilon)]}\right) \;, \qquad
\label{eq:Zjet-1}
\end{eqnarray}
with  $\bald{p}_{Q} = \bald{p}_{\rm(jet)} (1 -  ( 1 - f_{q,g} ) \epsilon)$.
Eq.~(\ref{eq:Zjet-1}) implies that the observed tagged jet cross section in A+A reactions is a probabilistic superposition of
cross sections for jets of higher initial transverse energy. This excess energy is then redistributed outside of the jet due to strong
final-state interactions. Here $d\sigma^{q,g}/{d^2 \bald{p}_{(Z)} d^2 \bald{p}_{\rm(jet)} }$ are the differential cross sections for away-side quark and gluon jets, respectively. 

In some experiments  the $Z^0/\gamma^*$+jet final-state channel is measured without placing restrictions on the momentum of the vector boson~\cite{Abazov:2008ez}. In this case, we can integrate over $\bald{p}_{(Z)}$ in Eq.~(\ref{eq:Zjet-1}) and derive:
\begin{eqnarray}
\frac{d\sigma^{AA} }{ d^2 \bald{p}_Q} =
\sum_{q,g} \int d \epsilon \; P_{q,g}(\epsilon) \frac{1}{[1 - (1-f_{q,g}(\omega_{\min},R))\epsilon]^2} 
 \frac{d\sigma^{q,g}}{ d^2 \bald{p}_{\rm(jet)}}  \left(\frac{\bald{p}_{Q}}
{[1 - (1-f_{q,g})\epsilon)]}\right) \;. \qquad
\label{eq:Zjet-2}
\end{eqnarray}
We notice that the integrated tagged jet cross section ${d\sigma^{AA} }/{ d^2 \bald{p}_Q} $ usually unravels the same physics as the suppression of the cross section in inclusive jet production discussed in Sec.~\ref{sec:jcs}. 

To demonstrate nuclear effects for tagged jet cross sections in HIC we define
 $I_{AA}^{\rm jet}$ for tagged jet  as:
\begin{equation}
I_{AA}^{\rm jet}(R,\omega_{\min}) = \frac{1}{\langle  T_{AA} \rangle \sigma^{\rm geo}_{AA}  } \frac{d\sigma_{AA}}{d p_{T\,(Z)}d p_{T\,(Q)}}
\Bigg / \frac{d\sigma_{pp}}{d p_{T\,(Z)}d p_{T\,(\rm jet)}} \; .
\label{iaa}
\end{equation}
While at LO $I_{AA}^{\rm jet}$ is limited to the $Z^0/\gamma^*$ trigger momentum range,
at NLO it will show rich features as we will discuss below.

A numerical simulation of $Z^0$/$\gamma^*$ tagged jet productions in p+p and Pb+Pb with $\sqrt{s_{NN}}=4$~TeV
for a cone size $R=0.2$ is provided in the right panel of Fig.~\ref{fig:Zjet} .  It is found because
part of the jet energy is redistributed outside of the jet cone the $p_T$ spectrum of tagged jet production in HIC  is downshifted
toward smaller transverse momenta. More interestingly one can observe a sharp transition of  $I_{AA}^{\rm jet}$ from tagged jet enhancement below $p_{T\, (Z)}$ to tagged jet suppression above $p_{T\, (Z)}$.
This striking feature provides a unique prediction of jet quenching for tagged jets and will be experimental evidence for strong final-state interactions and parton energy loss in the QGP. Please note that while the same energy redistribution occurs for inclusive jet production
discussed in Sec.~\ref{sec:jcs}, the monotonically falling spectrum prevents the observation of such inclusive jet enhancement. 

\section{Non-perturbative effects in jet productions}
\label{sec:NP}

%%%%%%%*********** Figure for non-perturbative **************
%\vspace{0.3in}
\begin{figure}[t]
\begin{center}
%\vspace*{-1.in}
\includegraphics[width=2.8in,height=2.3in,angle=0]{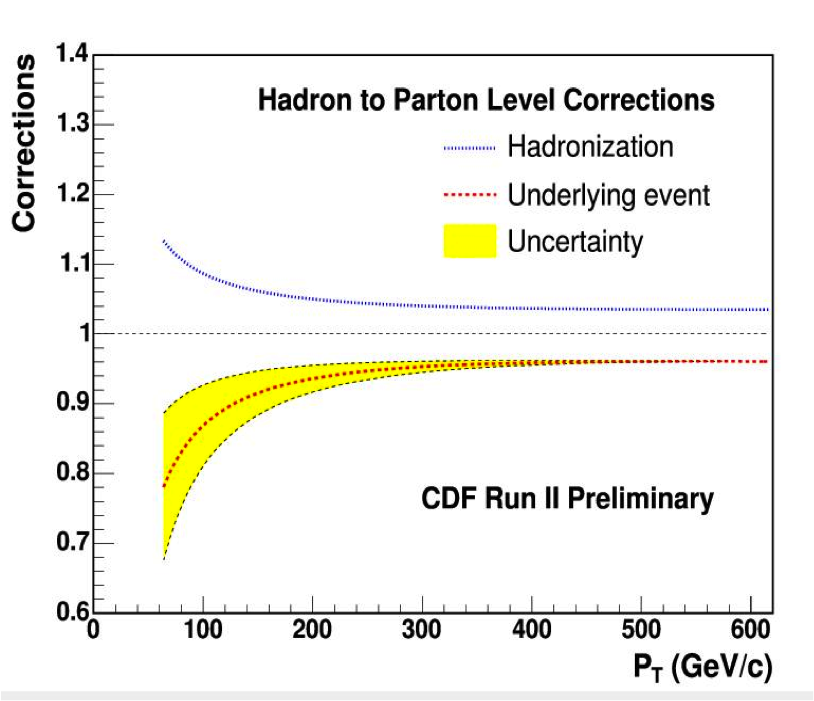} 
\hspace{0.6cm}
\includegraphics[width=2.8in,height=2.2in,angle=0]{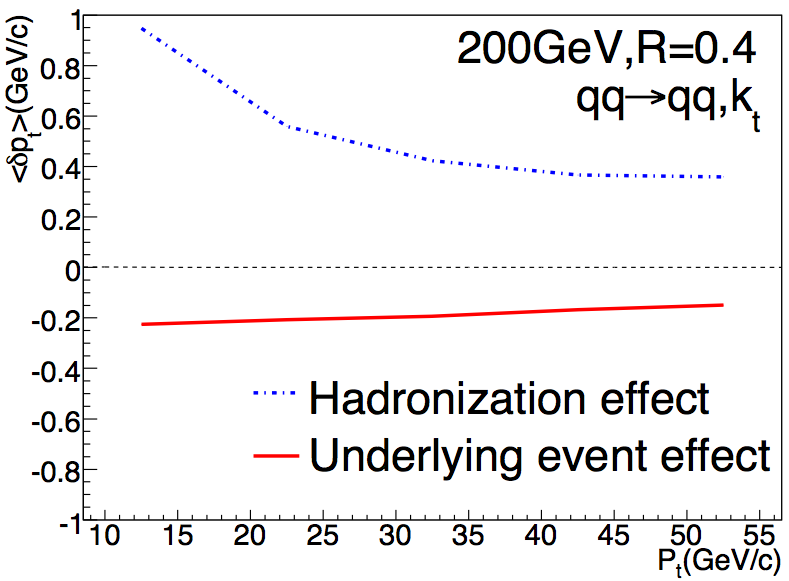} \vspace*{-.1in}
\caption{ Magnitude of the hadronization and underlying event effects
 used to correct the inclusive jet cross section measured by CDF with jet radius
$R=0.7$ \cite{Campbell:2006wx} (left panel).  Modification of $p_T$ of jets due to non-perturbative effects 
 in $p+p$ collisions at RHIC with Pythia8.142 \cite{NonP} (right panel). Here for all plots the corrections are taken
 from hadron to parton level.}
\label{fig:NP}
\end{center}
\end{figure}
%%%%%%%%*************************************************************
%%%%%%%%%%%%--------------------------------%%%%%%%%%%%%%%%%%%%%

In perturbative QCD calculations of inclusive jet yield and tagged jet production, a  jet is the combination
of a few partons defined by jet algorithms and what we calculate is the parton-jet.  However, in experiments what we measured 
is the hadron-jet since the final-state particles are hadrons. For jet production at high transverse momentum in a hard process with  a large
momentum transfer, the difference between parton-jets and hadron-jets may be small due to local parton-hadron
duality (LPHD) \cite{EKS,Campbell:2006wx}. Nevertheless, to make a more reliable comparison between QCD theory
of jet productions and the jet measurements, the theoretical result of jet cross sections 
with pQCD should be corrected to hadron level, or on the contrary the experimental data should be corrected to parton level 
\cite{Campbell:2006wx, NonP,Salam:2009jx}.

To include these corrections two important non-perturbative effects should be taken into account:
the hadronization effect which results in the energy loss of a jet because when partons in a parton-jet hadronize, some hadrons may fall outside of the jet ( a "splashing-out" effect ); the underlying event effects which gives the energy gain of a jet due to interaction of beam-beam remnants not associated with the hard scattering 
( a "splash-in" effect ).  The good news is that the hadronization effect and the underlying event effect
go in opposite direction and their contributions will be partially cancelled.

The left panel of Fig.~\ref{fig:NP} gives the magnitude of non-perturbative effects used to correct
the $p_T$ spectra  with a cone size $R=0.7$ for $p+{\bar p}$ collisions with $\sqrt{s_{NN}}=1960$~GeV 
by CDF at Tevatron. We find that when  $p_T>200$~GeV the fractional corrections due to non-perturbative effects
are within few percent.  However, this observation holds true only for this special case and corrections from 
the non-perturbative effects may vary with the colliding energy $\sqrt{s_{NN}}$,  jet transverse momenta $p_T$
and the jet radius $R$ \cite{Dasgupta:2007wa, NonP}. At RHIC with colliding energy $200$~GeV the measured jet does not have a very large $p_T$, 
and with small jet radii ($R=0.2,\ 0.4$  for measurements in Fig.~\ref{fig:jet@RHIC} )
the contributions from non-perturbative effects may be considerable.
In the right panel of Fig.~\ref{fig:NP} we show a numerical simulation of the jet $p_T$ shifting \cite{NonP} to inclusive jet production 
due to the hadronization effect and the underlying event
effects  in $p+p$ collisions at RHIC with Pythia8.142 \cite{pythia} . It could be seen that when $p_T \sim 12$~GeV the relative shifting ${\delta p_t }/p_T$ due to hadronization
could be about $8\%$, and its effect will be significantly amplified by the steeply falling $p_T$ spectrum of jets when we studying the jet cross section. Thus it is important to take into account
corrections of non-perturbative effects to jet productions in HIC at RHIC and see how jet spectra will be modified 
by these corrections \cite{NonP}.

\vspace*{1.0cm}

{ \bf Acknowledgments:}  The work  is finished in collaboration with I.~Vitev, S. Wicks and B. Neufeld.  
I thank X.~N.~Wang, J.~W.~Qiu and G.~Soyez for illuminating discussions. This research is  supported 
by the Ministry of Education of China with the Program NCET-09-0411,
by National Natural Science Foundation of China with Project No. 11075062,
and CCNU with Project No. CCNU09A02001.



%% References


%\bibliographystyle{elsarticle-num}



\end{document}